\documentclass[a4paper,11pt]{article}
\usepackage[utf8]{inputenc}
\usepackage{pos}
\usepackage{natbib}
\title{Initial State Summary of Hard Probes 2020}
\author*[a]{Bj\"orn Schenke}
\affiliation[a]{Physics Department, Brookhaven National Laboratory, Bldg. 510A, Upton, NY 11973, USA}

\emailAdd{bschenke@bnl.gov}

\abstract{The description of the initial state of heavy ion collisions, which covers the description of the incoming nuclei, the initial hard and soft interactions, the resulting spatial geometry of the produced matter, as well as the dynamic approach to a medium well described by hydrodynamics, has important consequences for the study of hard and electromagnetic probes. I will review new developments presented at Hard Probes 2020 that have an impact on these aspects of our understanding of the initial state of heavy ion and smaller system collisions.}

\FullConference{%
  HardProbes2020\\
  1-6 June 2020\\
  Austin, Texas}

\begin{document}
\maketitle

%\section{Introduction}
\section{Initial geometry}

It has been established that the azimuthal momentum anisotropy of produced hadrons in heavy ion collisions originates from the (hydrodynamic) response of the produced medium to the initial geometry in the plane transverse to the beamline \cite{Gale:2013da}. Understanding the precise geometry and its fluctuations over a wide range of collision systems and energies is thus crucial for describing the bulk observables in heavy ion collisions. 

A powerful new observable that correlates the elliptic anisotropy $v_2$ and the mean transverse momentum was suggested in \cite{Bozek:2016yoj}, and measurements by the ATLAS Collaboration \cite{Aad:2019fgl} were presented. The correlator is sensitive to the details of the initial state model and dominated by the initial geometry \cite{Bozek:2016yoj, Schenke:2020uqq} in all systems, as long as $dN_{\rm ch}/d\eta>10$ . For smaller multiplicities initial momentum anisotropies will begin to dominate the final correlator, resulting in a characteristic sign change as a function of multiplicity \cite{Giacalone:2020byk}. Experiments have begun to investigate this, but are challenged by the presence of ``non-flow'' contributions at small multiplicities.

Other observables that can distinguish different models, for example the difference between energy deposition $\sim \sqrt{T_A T_B}$ and $\sim T_A T_B$ (where $T_{A/B}$ are the position dependent target and projectile thickness functions), include the ratio of the four and two particle cumulants $v_n\{4\}/v_n\{2\}$ \cite{Carzon:2020ohg}. The prescription $\sim \sqrt{T_A T_B}$ is preferred by the Trento model \cite{Moreland:2014oya}, which does not include the possibility of $\sim T_A T_B$ scaling, which is predicted by the color glass condensate (CGC) effective field theory (EFT) / glasma \cite{McLerran:1993ni} %,McLerran:1993ka} 
and AdS/CFT \cite{Romatschke:2017ejr}. 

We particularly need to better understand the rapidity dependence of the initial state. Even at the highest LHC energies, measurements of flow harmonics show a rather strong rapidity dependence \cite{Chatrchyan:2012ta,Adam:2016ows}, especially in more peripheral events. More complex observables can provide better constraints for 3+1 dimensional models, for example the longitudinal flow decorrelation, which was studied in Xe+Xe and Pb+Pb systems by the ATLAS Collaboration \cite{Aad:2020gfz}. This measurement revealed that hydrodynamic models that can describe the ratios of harmonic flow coefficients between Xe+Xe and Pb+Pb collisions, fail to describe most of the rapidity dependence characterized by $v_n$ rapidity decorrelations. New initial state models for the three dimensional structure of heavy ion collisions are being developed \cite{Shen:2017bsr,Shen:2020jwv}, and especially at low collision energies dynamical energy deposition is required, resulting in complex three dimensional fluctuating initial energy and net baryon density distributions \cite{Shen:2017bsr}. Detailed data as that described above will be very useful for constraining such models.

Finally, input that helps constrain the transverse geometry in small system collisions, such as p/d+A collisions, can be obtained from measurements of photoproduction of vector mesons, as was done in \cite{Mantysaari:2016ykx} using data from diffractive $J/\psi$ production in e+p collisions at HERA. New data on $J/\psi$ photoproduction in d+Au collisions was presented by the STAR Collaboration \cite{Tu:2020}. The computation of the coherent and incoherent cross section in the CGC framework is sensitive to the average shape and fluctutions of the deuteron, and the data seems to be able to constrain both \cite{Mantysaari:2019jhh}, and will hopefully also provide more insight into short range nucleon correlations \cite{Tu:2020ymk}.

\vspace{-0.4cm}
\section{Nuclear effects}
\vspace{-0.4cm}
To properly describe the initial state of heavy ion collisions and understand for example the background to the medium modification of jets, nuclear modifications of parton distribution functions (PDFs) need to be well understood. A range of observables can be used to constrain nuclear PDFs, including quarkonium, electroweak boson, light meson, as well as dijet production. 

$J/\psi$ production in p+p collisions can be well described within a combined CGC+non-relativistic QCD (NRQCD) (+Fixed Order+Next-to-Leading Log (FONLL) \cite{Cacciari:2012ny}) framework \cite{Ma:2014mri}. The nuclear modification observed in p+Pb collisions \cite{Acharya:2018kxc,Acharya:2020wwy} is also well described within the CGC EFT \cite{Ducloue:2016pqr,Albacete:2017qng}, and also using constrained nuclear PDFs \cite{Kusina:2017gkz,Albacete:2017qng} and coherent energy loss \cite{Arleo:2014oha,Albacete:2017qng}. What has not yet been fully described theoretically is the correlation between $J/\psi$ production at backward and forward rapidities and midrapidity charged hadron production in p+Pb collisions \cite{Acharya:2020giw}. It will be interesting to see what physics aspects are needed to describe the observed trends.

The nuclear dependence of $J/\psi$ production as a function of rapidity was studied by PHENIX at RHIC \cite{Acharya:2019zjt}. As expected, in p+Au collisions one finds stronger shadowing in the forward direction and stronger nuclear absorption in the backward direction compared to p+Al collisions. For collisions with a Au target (p+Au and $^3$He+Au), only nuclear PDFs are not sufficient to describe the measured $R_{AB}$ ratios. These types of measurements can be very useful to understand the dependence on both nuclear projectile and target. 

Electroweak boson production is also sensitive to nuclear effects, as demonstrated by ALICE for Z \cite{Acharya:2020puh} and W$^{\pm}$ production (preliminary data presented at this conference), and new forward measurements can be used as additional input to global nuclear PDF fits. Electroweak boson data has also been suggested to be used to calibrate the Glauber model \cite{Eskola:2020lee}: Assuming that EPPS16 correctly describes nuclear effects on the production of W and Z bosons, using a reduced nucleon-nucleon inelastic cross section of $41.5^{+16.2}_{-12.0}\,{\rm mb}$ compared to the usual $70\pm 5\,{\rm mb}$ at $\sqrt{s}=5\,{\rm TeV}$ leads to much better agreement with the $R_{\rm PbPb}$ for W and Z bosons measured at LHC \cite{Aad:2019sfe,Aad:2019lan}, implying nuclear shadowing of the nucleon-nucleon cross section itself. 

The nuclear modification of light mesons has now been measured out to $p_T$ of up to 200 GeV in 8.16 TeV p+Pb collisions by ALICE and presented at this conference. The results are consistent with theory calculations within the CGC EFT \cite{Lappi:2013zma} and coherent energy loss \cite{Arleo:2020hat}. Further insight can be gained from small system scan collisions at RHIC, where the $\pi^0$ $R_{p/d/^3He+Au}$ has been measured by PHENIX (\cite{Adler:2006wg} and preliminary data). Similar to p+Pb collisions at LHC, the centrality dependence observed is consistent with fluctuating proton sizes \cite{McGlinchey:2016ssj}.

Finally, dijet measurements in p+p and p+A collisions provide additional information on nuclear effects. For example, predictions for forward dijet angular correlations were presented within the CGC framework \cite{Wei:2020}. Because of Sudakov effects, which depend on the kinematics and can mask saturation effects, the ideal range in dijet transverse momenta to observe saturation effects, was identified to be 5 to 8 GeV (at forward rapidities) in $\sqrt{s}=8.8\,{\rm TeV}$ collisions. The CMS Collaboration presented measurements of the ratio of rapidity dependent dijet production in p+Pb over p+p collisions \cite{Sirunyan:2018qel}. This provided first evidence that apart from suppression at small $x$, the gluon PDF at large $x$ in Pb is strongly suppressed with respect to the PDF in unbound nucleons.

\vspace{-0.4cm}
\section{Theory progress at small x}
\vspace{-0.4cm}
Steady progress is being made in extending calculations within the CGC EFT to next-to-leading order (NLO). First phenomenological applications using the full NLO impact factor and the NLO Balitsky-Kovchegov (BK) equation \cite{Balitsky:2008zza,Iancu:2015vea,Lappi:2015fma,Lappi:2016fmu} to determine the $x$ dependence of deep inelastic scattering (DIS) observables have become available \cite{Beuf:2020dxl}. While the calculation does only include terms that are enhanced by large transverse logarithms, three different formulations of these evolution equations (which resum some or all of the transverse logarithms) yield little difference. Like at leading order (LO), there seems to be a significant nonperturbative contribution to the structure function for light quarks. Extension of the NLO calculation of the impact factor for heavy quarks, which would improve this situation, is still outstanding.

The BK equation can be viewed as the large $N_c$ limit of the JIMWLK equation \cite{JalilianMarian:1997dw,Iancu:2001ad}, which has also been extended to NLO \cite{Balitsky:2013fea,Kovner:2013ona}. At LO, it is known that finite $N_c$ corrections are small and the BK equation approximates the dipole evolution well \cite{Rummukainen:2003ns,Dumitru:2011vk}. At NLO, appearing six-point functions factorize into dipole operators at large $N_c$. Using the Gaussian approximation for these higher point correlators, a closed finite-$N_c$ BK equation at NLO was obtained \cite{Lappi:2020srm}. Again, it was shown that finite $N_c$ corrections are smaller than the naively expected $1/N_c^2\sim\mathcal{O}(10\%)$.

A complete NLO calculation of inclusive dijet+photon production in e+p/A collisions within the CGC EFT has recently been outlined in \cite{Roy:2018jxq,Roy:2019hwr,Roy:2019cux}. This involves in particular the calculation of the NLO impact factors. The calculation is powerful, as it includes many NLO processes at small $x$, such as inclusive dijet, inclusive photon+jet, and inclusive photon production, as well as fully inclusive DIS. It further carries information on LO $q\bar{q}g$ and $q\bar{q}g+\gamma$ production. Combining these results for the NLO impact factor with next-to-leading-log (NLL) resummed evolution kernels, an accuracy of order $\alpha_s^3\ln(1/x)$ can be achieved. Based on this, explicit calculations for prompt photon+jet production can be done at NLO. First LO results for this process, studying the azimuthal photon-jet correlations, which carry information on saturation, have recently been published in \cite{Kolbe:2020tlq}.

Further progress reported included extensions beyond the eikonal approximation: 1) in the calculation of quark-nucleus scattering in a light-front Hamiltonian approach \cite{Li:2020uhl}, and 2) by including large $x$ gluons in the target, which also aims at developing a unified picture for particle production at small and large $x$ \cite{Jalilian-Marian:2020dwf}. A generalized connection between transverse momentum dependent PDFs (TMDs) and the CGC EFT, using a new approach to TMDs at small $x$ employing transverse gauge link operators \cite{Boussarie:2020vzf} was presented, as well as the first computation of entanglement and ``ignorance'' entropies within the CGC EFT, the latter reflecting our inability to perform a complete set of measurements (sensitive to off-diagonal elements of the density matrix) \cite{Duan:2020jkz}.

\vspace{-0.4cm}
\section{Approach to equilibrium}
\vspace{-0.4cm}
Another area where a lot of progress has been made in the last couple of years is the study of the early time non-equlibrium evolution that is expected to result in a system that is well approximated by hydrodynamics \cite{Florkowski:2017olj,Berges:2020fwq}. The question of when hydrodynamics is applicable can be studied in both strong and weak coupling limits \cite{Kurkela:2019set}. In the strong coupling limit, described within AdS/CFT, causality requires that hydrodynamization occurs on time scales of order $\sim 1/T$, where $T$ is the temperature, as this is the distance from the horizon to the boundary, where the field theory lives. In the weak coupling limit, described by the Boltzmann equation and a relaxation time approximation (RTA) for the collision term, hydrodynamization is found to occur within $2\tau_0$, where $\tau_0$ is the initial time. The two limits behave somewhat differently, as in the RTA solutions approach the attractor before any finite order hydrodynamic expansion agrees with the attractor, while in AdS/CFT 2nd order hydrodynamics is usually a good approximation by the time the solutions approach the attractor.

Also chemical equilibration has been studied using the Boltzmann equation for quarks and gluons including $2\leftrightarrow 2$ and $1\leftrightarrow 2$ processes \cite{Kurkela:2018xxd,Kurkela:2018oqw}. It was found that chemical equilibration occurs after hydrodynamization but before local thermalization. Interestingly, chemical equilibration occurs for all systems with $dN_{\rm ch}/d\eta \gtrsim 100$, independent of the system size. Furter studies \cite{Du:2020} have found that chemical and kinetic equilibration for an expanding QGP with non-zero charge density occurs on roughly the same time scale as for zero charge density systems. Delayed chemical equilibration (relative to ``hydrodynamization'') can have an important effect on photon production and will affect photon anisotropic flow, as was explicitly shown in \cite{Gale:2020xlg} using a hybrid calculation employing IP-Glasma initial state \cite{Schenke:2012wb}, KoMPoST pre-equilibrium evolution \cite{Kurkela:2018vqr}, \textsc{Music} hydrodynamics \cite{Schenke:2010nt,Schenke:2010rr}, and UrQMD hadronic afterbuner \cite{Bass:1998ca}.

\vspace{-0.4cm}
\section{Conclusions}
\vspace{-0.4cm}
Our understanding of the initial state in heavy ion collisions and the early time dynamics is continuously improving. A lot of progress is being made in the first principles understanding of the nuclear wave function and nuclear and saturation effects, specifically the CGC EFT is being extended to NLO and calculations for a variety of processes at NLO are emerging. Furthermore, the initial geometry and its fluctuations are being better constrained, including the longitudinal direction, and the approach to a system that is described by hydrodynamics is being understood increasingly well. Among many other subfields of high energy nuclear physics, the study of hard and electromagnetic probes in heavy ion collisions will benefit immensly from these developments.

\vspace{-0.4cm}
\section*{Acknowledgments}
\vspace{-0.4cm}
B.P.S. is supported under DOE Contract No. DE-SC0012704.
\vspace{-0.4cm}

\bibliographystyle{JHEP} 
\bibliography{hp}

\end{document}